\documentclass[submission,copyright,creativecommons]{eptcs}
\providecommand{\event}{ThEdu 2017} 

\usepackage{amsmath}
\usepackage{graphicx,xcolor}
\usepackage{bussproofs}
\usepackage{xspace}
\usepackage{verbatim}

\usepackage{tikz,pgf}
\usetikzlibrary{shapes.geometric,shapes.arrows,chains,positioning,fit,shadows,decorations.pathreplacing,automata,shapes.misc,arrows}

\usepackage{ntheorem}

\newtheorem{theorem}{Theorem}
\newtheorem{proposition}[theorem]{Proposition}
\theoremstyle{plain}
\theorembodyfont{\normalfont}

\newtheorem{example}{Example}

\newcommand{\Var}{\ensuremath{\mathcal{V}}}

\newcommand{\fo}[1]{\textup{FOL}[#1]\xspace}

\newcommand{\imp}{\ensuremath{\rightarrow}}
\newcommand{\biimp}{\ensuremath{\leftrightarrow}}

\newcommand{\arity}[1]{\ensuremath{\mathit{ar}(#1)}}

\newcommand{\yields}{\ensuremath{\Longrightarrow}}
\newcommand{\sem}[3]{\ensuremath{[\![ #1 ]\!]^{#2}_{#3}}}

\newcommand{\derives}{\;\Longrightarrow\;}

\newcommand{\AndRightRule}{\ensuremath{\mathtt{(\wedge_R)}}}
\newcommand{\AndLeftRule}{\ensuremath{\mathtt{(\wedge_L)}}}
\newcommand{\OrRightRule}{\ensuremath{\mathtt{(\vee_R)}}}
\newcommand{\OrLeftRule}{\ensuremath{\mathtt{(\vee_L)}}}
\newcommand{\NegRightRule}{\ensuremath{\mathtt{(\neg_R)}}}
\newcommand{\NegLeftRule}{\ensuremath{\mathtt{(\neg_L)}}}
\newcommand{\ImpRightRule}{\ensuremath{\mathtt{(\imp_R)}}}
\newcommand{\ImpLeftRule}{\ensuremath{\mathtt{(\imp_L)}}}

\newcommand{\AxiomRule}{\ensuremath{\mathtt{(Ax)}}}
\newcommand{\ForallLeftRule}{\ensuremath{\mathtt{(\forall_L)}}}
\newcommand{\ForallRightRule}{\ensuremath{\mathtt{(\forall_R)}}}
\newcommand{\ExistsLeftRule}{\ensuremath{\mathtt{(\exists_L)}}}
\newcommand{\ExistsRightRule}{\ensuremath{\mathtt{(\exists_R)}}}
\newcommand{\ContrLeftRule}{\ensuremath{\mathtt{(Contr_L)}}}
\newcommand{\ContrRightRule}{\ensuremath{\mathtt{(Contr_R)}}}
\newcommand{\SubstLeftRule}{\ensuremath{\mathtt{(Subst_L)}}}
\newcommand{\SubstRightRule}{\ensuremath{\mathtt{(Subst_R)}}}
\newcommand{\EqLeftRule}{\ensuremath{\mathtt{(Eq_L)}}}
\newcommand{\EqRightRule}{\ensuremath{\mathtt{(Eq_R)}}}

\renewcommand{\phi}{\varphi}

\newcounter{TODOcounter}

\title{The Sequent Calculus Trainer with Automated Reasoning \\ -- Helping Students to Find Proofs} 

\author{Arno Ehle \qquad\qquad Norbert Hundeshagen \quad\quad Martin Lange
	\institute{School of Electrical Engineering and Computer Science \\
			University of Kassel, Germany}
	\email{\quad post@arnoehle.de \quad\qquad hundeshagen@uni-kassel.de \quad\quad mlange@uni-kassel.de}
}
\def\titlerunning{The Sequent Calculus Trainer with Automated Reasoning}
\def\authorrunning{A.~Ehle, N.~Hundeshagen \& M.~Lange}

\begin{document}

\maketitle

\begin{abstract}
The sequent calculus is a formalism for proving validity of statements formulated in First-Order Logic. It is 
routinely used in computer science modules on mathematical logic. Formal proofs in the sequent calculus are
finite trees obtained by successively applying proof rules to formulas, thus simplifying them step-by-step.

Students often struggle with the mathematical formalities and the level of abstraction that topics like formal
logic and formal proofs involve. The difficulties can be categorised as syntactic or semantic. On the syntactic
level, students need to understand what a correctly formed proof is, how rules can be applied (on paper for
instance) without leaving the mathematical framework of the sequent calculus, and so on. Beyond this, on the 
semantic level, students need to acquire strategies that let them find the right proof. 
  
The Sequent Calculus Trainer is a tool that is designed to aid students in learning the techniques of proving given statements formally. 
In this paper we describe 
the didactical motivation behind the tool and the techniques used to address issues on the syntactic as well as on the semantic level.
\end{abstract}


\section{Introduction}
\label{sect:introduction}

Complaints among computer science students about the perceived difficulty and toughness of theoretical modules in their curricula are
ubiquitous. This can typically be attributed to the comparably more formal content of those courses, which require a deeper understanding
of the taught material in order to successfully solve exercises and ultimately pass exams.

This paper addresses one particular such issue, occurring in courses on formal logic. Most university curricula in computer science include 
such a course as a basic module, as part of either the mathematical foundations or the theoretical computer science strand. In fact, 
recommendations on the content of computer science courses usually include mathematical logic in the form of a stand-alone module or as part of a more
general module on discrete structures, c.f.\ \cite{ACMIEEERec13,GIEmpf16}, and this should include the first-order predicate calculus besides
basic propositional logic. 

Teaching mathematical logic typically includes the study of a syntactic proof system like natural deduction \cite{Gentzen35a,Jaskowski:1934},
resolution \cite{DP60,Robinson65}, Hilbert proof systems or the sequent calculus \cite{Gentzen35a,Szabo1969}. Some typical basic-level exercises
ask for a formal proof of a given formula of first-order logic. One of the greatest challenges for the students there is to understand the
connection between semantic and syntactic deduction. Even if the student has an intuitive idea of why the given formula is valid, turning
this intuition into a formal proof still requires them to have obtained the understanding of this deep logical connection. So finding a proof 
for a given formula requires the students to have acquired the ability to reason using both syntactic and semantic tools -- the former for a
rigorous formulation of proof steps and the latter for a deeper understanding of the underlying mathematical structures involved (c.f.\ 
\cite{Weber2004}). 

It is worth noting that this connection between semantic and syntactic reasoning is enabled mathematically through the existence of 
completeness theorems for the underlying calculi, in the sense of Hilberts programm, G\"{o}del's Completeness Theorem \cite{Goedel:1930}, 
Gentzen's Hauptsatz \cite{Gentzen35a,Szabo1969} and so on. One of the main reasons for teaching completeness as a central concept is to form this
connection in students' minds. This is beneficial not only for learning formal logic but for the computer science education in general
given that the essence of finding algorithmic solutions to any kind of problem lies in a syntactical characterisation of this 
problem for otherwise computers would not be able to carry out the solution by any means of symbolic manipulation of variable values,
memory contents, etc.

The Bachelor's curriculum for computer science at the University of Kassel contains a mandatory 2nd-year course on formal logic, which focuses on
the model and proof theory of first-order logic with equality. Propositional logic is presented as a true fragment of first-order logic.
The course has been rigorously shaped with the aim of improving learning outcomes and therefore reducing failure rates. A central point of 
its organisation is the use of constructivistic learning theory, i.e.\ students are expected to learn formal logic in a highly self-organised
and self-regulated way. This includes machanisms like electronic feedback systems, tool support, highly structured learning material and -- 
to a smaller degree -- the use of the inverted-classroom model (c.f.\ \cite{LPT2000}). This model focuses on learning, literally, as a 
self-organised activity; consequently, the course engages students with methods and tools to assist and self-assess the use of formal logic 
and the calculi taught with them. 

One of these tools, developed for such purposes, is the \emph{Sequent Calculus Trainer} (SCT). In  \cite{EHL:TTL15} a first version has 
been introduced, which solely focused on reducing mistakes made by students on the syntactic level. Empirical data in the form of exam 
results suggest that this earlier version of SCT indeed helps students to master the challenges of this level. Namely, a reduction in
syntactical mistakes in corresponding exam papers could be noticed, and it could also be linked to the use of the SCT tool.

However, this feature alone is not enough to train students adequately; the semantical level still needs to be achieved. In other words,
understanding what a correct proof is, is only the first step in finding one. Consequently, the Sequent Calculus Trainer has been extended
to tackle the problem of helping students to master the semantical level as well. It uses a simple feedback system which can give the
user a hint about how to construct a proof, first of all by directly issuing a warning when a bad step has been taken (i.e.\ a rule has
been applied by the user such that the resulting subgoals have been identified to be unprovable), but also by being able to make 
suggestions about which rule how to apply next in order to get closer to finishing the construction of the formal proof. 

The aim of this paper is to present the background and internal technology of the Sequent Calculus Trainer, now 
capable of assisting students to find the right proof and thus also addressing the aforementioned semantic level of correct reasoning.

In the following the word calculus refers to the sequent calculus, as presented in Section\ \ref{sec:seq_calc}.


\section{The Sequent Calculus for First-Order Logic with Equality}\label{sec:seq_calc}

This section recalls the definition of the syntax and semantics of First-Order Logic with Equality over uninterpreted function symbols
and a formal proof system for validity known as the Sequent Calculus.
 
\subsection{First-Order Logic with Equality}

A \emph{signature} is a list $\tau = \langle R_1,\ldots,R_n,f_1,\ldots,f_m \rangle$ of \emph{relation symbols} $R_i$ and 
\emph{function symbols} $f_i$. Each of these implicitly has an \emph{arity} denoted by $\arity{R}$, resp.\ $\arity{f}$.

Let $\Var = \{x,y,z,\ldots\}$ be a countably infinite set of (first-order) variables. \emph{Formulas} $\varphi,\psi$ and \emph{terms} 
$t_1,t_2,\ldots$ of First-Order Logic with Equality over $\tau$, $\fo{=,\tau}$ for short, are given by the following grammar. 
\begin{align*}
\varphi,\psi \enspace &::= \enspace R(t_1,t_2,\ldots, t_{\arity{R}}) \mid t_i = t_j \mid \varphi \wedge \psi \mid 
  \varphi \vee \psi \mid \neg\varphi \mid \varphi \imp \psi \mid \varphi \biimp \psi \mid \exists x\,\varphi \mid \forall x\,\varphi \\
t_1,t_2,\ldots \enspace &::= \enspace x \mid f(t_1,t_2,\ldots, t_{\arity{f}})
\end{align*}
where $x \in \Var$, $R$ is a relation symbol in $\tau$ and $f$ is a function symbol in $\tau$. Note that function symbols of arity
$0$ are terms as well, called \emph{constants}.

Terms and formulas are interpreted in \emph{$\tau$-structures} $\mathcal{A}$, consisting of a set $U$ -- called the \emph{universe},
and,
\begin{itemize}
\item for each $i=1,\ldots,n$, a relation $R^{\mathcal{A}}$ over $U$ of arity $\arity{R}$, and 
\item for each $i=1,\ldots,m$, a total function $f^{\mathcal{A}}$ on $U$ of arity $\arity{f}$.
\end{itemize}
A \emph{$\tau$-interpretation} is a pair $(\mathcal{A},\vartheta)$ consisting of a $\tau$-structure $\mathcal{A}$ with a universe
$U$ and a variable interpretation $\vartheta: \Var \to U$.

Terms denote elements of $\tau$-structures; the value of term $t$ under a $\tau$-interpretation $(\mathcal{A},\vartheta)$ is 
denoted by $\sem{t}{\mathcal{A}}{\vartheta}$ and is obtained by successively applying the functions corresponding to the symbols
in the term, starting with those elements pointed to by the variable interpretation:
\begin{displaymath}
\sem{x}{\mathcal{A}}{\vartheta} := \vartheta(x) \quad, \quad 
\sem{f(t_1,\ldots,t_m)}{\mathcal{A}}{\vartheta} := 
f^{\mathcal{A}}(\sem{t_1}{\mathcal{A}}{\vartheta},\ldots,\sem{t_m}{\mathcal{A}}{\vartheta})
\end{displaymath}

The satisfaction of a $\fo{=,\tau}$-formula $\varphi$ by a $\tau$-interpretation $(\mathcal{A},\vartheta)$ is denoted by
$\mathcal{A},\vartheta \models \varphi$ and is inductively explained as follows.
\begin{displaymath}
\begin{aligned}
\mathcal{A},\vartheta &\models R(t_1,\ldots,t_n) \enspace &&\Leftrightarrow \quad 
  (\sem{t_1}{\mathcal{A}}{\vartheta},\ldots,\sem{t_n}{\mathcal{A}}{\vartheta}) \in R^{\mathcal{A}} \\
\mathcal{A},\vartheta &\models t = t' \enspace &&\Leftrightarrow \quad \sem{t}{\mathcal{A}}{\vartheta} = \sem{t'}{\mathcal{A}}{\vartheta} \\
\mathcal{A},\vartheta &\models \varphi \wedge \psi \enspace &&\Leftrightarrow \quad 
  \mathcal{A},\vartheta \models \varphi \text{ and } \mathcal{A},\vartheta \models \psi \\
\mathcal{A},\vartheta &\models \varphi \vee \psi \enspace &&\Leftrightarrow \quad 
  \mathcal{A},\vartheta \models \varphi \text{ or } \mathcal{A},\vartheta \models \psi \\
\mathcal{A},\vartheta &\models \neg\varphi \enspace &&\Leftrightarrow \quad \mathcal{A},\vartheta \not\models \\
\mathcal{A},\vartheta &\models \varphi \imp \psi \enspace &&\Leftrightarrow \quad 
  \mathcal{A},\vartheta \models \varphi \text{ implies } \mathcal{A},\vartheta \models \psi \\
\mathcal{A},\vartheta &\models \varphi \biimp \psi \enspace &&\Leftrightarrow \quad 
  \mathcal{A},\vartheta \models \varphi \text{ iff } \mathcal{A},\vartheta \models \psi \\
\mathcal{A},\vartheta &\models \exists x\,\varphi \enspace &&\Leftrightarrow \quad 
  \text{there is } u \in U \text{ such that } \mathcal{A},\vartheta[x \mapsto u] \models \varphi \\
\mathcal{A},\vartheta &\models \forall x\,\varphi \enspace &&\Leftrightarrow \quad 
  \text{for all } u \in U \text{ we have } \mathcal{A},\vartheta[x \mapsto u] \models \varphi
\end{aligned}
\end{displaymath}
where $\vartheta[x \mapsto u]$ denotes the update of $\vartheta$ at position $x$ with value $u$.

A $\fo{=,\tau}$-formula $\varphi$ is \emph{valid} iff for all $\tau$-interpretations $(\mathcal{A},\vartheta)$ we have 
$\mathcal{A},\vartheta \models \varphi$. Examples of valid formulas include
\begin{displaymath}
\forall y\, R(z,y) \imp \forall y\,\exists x\,R(x,y)\enspace, \quad
\forall x\, f(x) = x \to \forall x\, f(f(x))=x\enspace,\quad 
\exists x\,(\forall y\, \mathit{Drinks}(y) \imp \mathit{Drinks}(x))
\end{displaymath}

In the following we will simply speak of First-Order Logic, FOL in short, when we mean First-Order Logic with Equality over a 
particular signature $\tau$ which is derivable from the context. We further assume, that all formulas are \textit{sentences}, i.e. all variables are bound by a quantifier. Note that models
for sentences can be given as a $\tau$-structure alone, i.e.\ no variable assignment is needed.

\subsection{Sequents and Validity}

The Sequent Calculus is a formal system with which one can derive the validity of a formula by purely symbolic formula manipulation.
The notion of validity of a formula is suitably generalised in order to enable the application of simple logical principles in the form
of proof rules. The basic data structure for this purpose is that of a \emph{sequent} -- a pair of finite multisets of formulas written
$\Gamma \derives \Delta$. The left part $\Gamma$ is called the \emph{antecedent}; the right part $\Delta$ is called the \emph{succedent}
of the sequent.

Such a sequent is \emph{valid} if the formula $(\bigwedge \Gamma) \imp \bigvee \Delta$ is valid. Hence, in a sequent
$\varphi_1,\ldots,\varphi_n \derives \psi_1,\ldots,\psi_m$, the comma separating different elements of the (multi-)set is 
interpreted as a conjunction in the antecedent and as a disjunction in the succedent. Note that validity of finite sequents does 
indeed generalise validity of formulas since the formula $\varphi$ is valid iff the sequent $\emptyset \derives \varphi$ is valid.
Likewise, validity of finite sequents can be expressed as validity of a formula as it is done in the definition here.

A sequent $\Gamma \derives \Delta$ of $\tau$-formulas is consequently \emph{invalid} if there is a $\tau$-interpretation
$(\mathcal{A},\vartheta)$ that fulfils all formulas of the antecedent but none of the succedent, i.e.\ 
$\mathcal{A},\vartheta \models \varphi$ for all $\varphi \in \Gamma$ and $\mathcal{A},\vartheta \not\models \psi$ for all 
$\psi \in \Delta$. Such an interpretation is also called a \emph{countermodel} for $\Gamma \derives \Delta$.

\subsection{Formal Proofs}\label{subsec:formal_proofs}

The Sequent Calculus is a formal proof system that characterises the semantic notion of validity of sequents (and therefore validity of 
formulas) through the existence of a purely syntactic object, namely a \emph{formal proof} for a sequent $\Gamma \derives \Delta$ under
consideration. Such a formal proof is a finite tree whose nodes are labeled with sequents, such that
\begin{itemize}
\item the tree's root is labeled with $\Gamma \derives \Delta$, and
\item the labels at each node, together with the labels on its children, form a substitution instance of one of the proof rules depicted in 
      Figures~\ref{fig:rulesbool}--\ref{fig:axioms}.
\end{itemize}
We let proof trees grow upwards, i.e.\ the sequent under consideration is shown at the bottom of the tree.
 
Rules are of the form
\begin{displaymath}
    (N) \enspace
	\begin{array}{c}
	\Gamma_1 \derives \Delta_1 \quad \ldots \quad \Gamma_n \derives \Delta_n \\ \hline
	\Gamma \derives \Delta
	\end{array}
\end{displaymath}
for some $n \in \{0,1,2\}$. $(N)$ is the rule's name simply used to identify it when reasoning about proofs. Each $\Gamma_i \derives \Delta_i$ 
is called a \emph{premiss} of the rule, and $\Gamma \derives \Delta$ is called the \emph{conclusion}. Note that some rules have no 
premisses -- they are called \emph{axioms} -- and they are the only rules that can be used to close a branch of a proof tree. 

\begin{figure}[t]
	\begin{displaymath}
	\AndLeftRule\enspace
	\begin{array}{c}
	\Gamma, \varphi, \psi \Longrightarrow \Delta \\ \hline
	\Gamma, \varphi \wedge \psi \Longrightarrow \Delta
	\end{array}
	\qquad
	%
	\AndRightRule\enspace
	\begin{array}{c}
	\Gamma \Longrightarrow \varphi, \Delta \qquad \Gamma \Longrightarrow \psi, \Delta \\ \hline
	\Gamma \Longrightarrow \varphi \wedge \psi, \Delta
	\end{array}
	\qquad
	%
	\NegLeftRule\enspace
	\begin{array}{c}
	\Gamma \Longrightarrow \varphi, \Delta \\ \hline
	\Gamma, \neg\varphi \Longrightarrow \Delta
	\end{array}
	\end{displaymath}
	
	\begin{displaymath}
	%
	\OrLeftRule\enspace
	\begin{array}{c}
	\Gamma, \varphi \Longrightarrow \Delta \qquad \Gamma, \psi \Longrightarrow \Delta \\ \hline
	\Gamma, \varphi \vee \psi \Longrightarrow \Delta
	\end{array}
	\qquad
	%
	\OrRightRule\enspace
	\begin{array}{c}
	\Gamma \Longrightarrow \varphi, \psi, \Delta \\ \hline
	\Gamma \Longrightarrow \varphi \vee \psi, \Delta
	\end{array}
	\qquad
	%
	\NegRightRule\enspace
	\begin{array}{c}
	\Gamma, \varphi \Longrightarrow \Delta \\ \hline
	\Gamma \Longrightarrow \neg\varphi, \Delta
	\end{array}
	\end{displaymath}
	
	\begin{displaymath}
	%
	\ImpLeftRule\enspace
	\begin{array}{c}
	\Gamma, \psi \Longrightarrow \Delta \qquad \Gamma \Longrightarrow \varphi, \Delta \\ \hline
	\Gamma, \varphi \to \psi \Longrightarrow \Delta
	\end{array}
	\qquad\enspace
	%
	\ImpRightRule\enspace
	\begin{array}{c}
	\Gamma, \varphi \Longrightarrow \psi, \Delta \\ \hline
	\Gamma \Longrightarrow \varphi \to \psi, \Delta
	\end{array}
	\end{displaymath}
	\caption{The proof rules for Boolean operators.}
	\label{fig:rulesbool} 
\end{figure}

Proof search in the sequent calculus thus amounts to the construction of a finite tree, starting with the sequent to be proved, then
selecting rules and applying them to the current sequent. This may create further proof obligations in the form of one or two premisses
of the currently applied rule, which then need to be handled in the same way until all created branches of the proof tree are closed
by the application of axioms.
 
The general intuition behind the format of the rules is the following. There is one rule for each potential occurrence of a logical 
operator at top-level in the antecedent and the succedent. These rules then try to simplify the sequent at hand by eliminating this
logical connector. This is at least true for the rules shown in Figure~\ref{fig:rulesbool}, which handle Boolean operators. Note, for instance
how the implicitly understood meaning of a sequent -- the antecedent is interpreted conjunctively, the succedent disjunctively -- is used
by rules $\AndLeftRule$ and $\OrRightRule$ to eliminate the occurrence of a conjunction in the antecedent or a disjunction in the 
succedent. 

Conversely, a conjunction in a succedent can be handled using rule $\AndRightRule$ which creates two ``smaller'' sequents to be proved.
This rule incorporates the distribution law for conjunctions and disjunctions by turning a statement about a disjunction including a
conjunction into two statements about disjunctions that both need to be fulfilled.

\begin{figure}[t]	%
	\begin{displaymath}
	%
	\ExistsLeftRule \enspace 
	\begin{array}{c}
	\Gamma, \varphi[c/x] \Longrightarrow \Delta \\ \hline
	\Gamma, \exists x\;\varphi \Longrightarrow \Delta
	\end{array}
	\enspace c \text{ fresh}
	\qquad
	%
	\ExistsRightRule \enspace
	\begin{array}{c}
	\Gamma \Longrightarrow \varphi[t/x], \Delta \\ \hline
	\Gamma \Longrightarrow \exists x\;\varphi, \Delta
	\end{array}
	\enspace t \text{ ground}
	\qquad
	%
	\ContrLeftRule\enspace
	\begin{array}{c}
	\Gamma, \varphi, \varphi \Longrightarrow \Delta \\ \hline
	\Gamma, \varphi \Longrightarrow \Delta
	\end{array}
	\end{displaymath}
	
	\begin{displaymath}
	%
	\ForallLeftRule \enspace
	\begin{array}{c}
	\Gamma, \varphi[t/x] \Longrightarrow \Delta \\ \hline
	\Gamma, \forall x\;\varphi \Longrightarrow \Delta
	\end{array}
	\enspace t \text{ ground}
	\qquad
	%
	\ForallRightRule \enspace
	\begin{array}{c}
	\Gamma \Longrightarrow \varphi[c/x], \Delta \\ \hline
	\Gamma \Longrightarrow \exists x\;\varphi, \Delta
	\end{array} 
	\enspace c \text{ fresh}
	\qquad
	%
	\ContrRightRule\enspace
	\begin{array}{c}
	\Gamma \Longrightarrow \varphi,\varphi,\Delta \\ \hline
	\Gamma \Longrightarrow \varphi, \Delta
	\end{array}
	\end{displaymath}

	\begin{displaymath}
	%
	\SubstLeftRule \enspace
	\begin{array}{c}
	\Gamma, \varphi[s'/x] \Longrightarrow \Delta \\ \hline
	\Gamma, s=s', \varphi[s/x] \Longrightarrow \Delta
	\end{array}
	\qquad
	%
	\SubstRightRule \enspace
	\begin{array}{c}
	\Gamma \Longrightarrow \varphi[s'/x], \Delta \\ \hline
	\Gamma, s=s' \Longrightarrow \varphi[s/x], \Delta
	\end{array}
	\qquad
	%
	\EqLeftRule \enspace
	\begin{array}{c}
	\Gamma, s=s \Longrightarrow \Delta \\ \hline
	\Gamma \Longrightarrow \Delta
	\end{array}
	\end{displaymath}

	\caption{The proof rules for quantifiers and equalities.}
	\label{fig:rulesquant} 
\end{figure}

\begin{figure}[t]
	\begin{displaymath}
	%
	\AxiomRule \enspace
	\begin{array}{c}
	\qquad \\ \hline
	\Gamma, \varphi \Longrightarrow \varphi, \Delta
	\end{array}
	\quad\enspace
	%
	\EqRightRule \enspace
	\begin{array}{c}
	\qquad \\ \hline
	\Gamma \Longrightarrow s=s, \Delta
	\end{array}
	\end{displaymath}

    \caption{The axioms.}
    \label{fig:axioms}    
\end{figure}

Quantifiers are handled in a similar way, in that they are also eliminated in an application of the respective rules as they are shown
in Figure~\ref{fig:rulesquant}. It is important to obey the requirements that in an application of rule $\ExistsLeftRule$ or 
$\ForallRightRule$, the respective variable is replaced by a fresh constant symbol $c$ which does not occur anywhere in the sequent of
the conclusion. Such constants are also sometimes referred to as \emph{Skolem constants}. Moreover, rules $\ForallLeftRule$ and 
$\ExistsRightRule$ require the replacement of the respective variable by a ground term, not an arbitrary term. It is not hard to construct
 examples of non-valid formulas which would be provable without these requirements.

On the other hand, one can equally construct examples of valid formulas which are not provable with those rules alone. Sometimes, the
formal proof of a statement from some assumption requires some assumption to be used more than once, in particular when it is a 
universally quantified formula. This is what rule $\ContrLeftRule$ is for. Likewise -- but perhaps less obviously -- one may need several 
copies of an existentially quantified formula in the succedent of a sequent which can be obtained with rule $\ContrRightRule$.

At last, equality possesses some fundamental principles which require three more rules which break the symmetry that is present in the
set of rules introduced so far. $\SubstLeftRule$ and $\SubstRightRule$ can be used to replace a term by another, provided that their
equality is assumed, i.e.\ is part of the sequent's antecedent. Note that these rules are only applicable if the replacement of term
$s$ by term $s'$ in $\varphi$ can be carried out harmlessly, i.e.\ no new variable bindings are created in this way. Finally, equality is
reflexive and it is therefore always possible to assume that some term equals itself using rule $\EqLeftRule$.

At last, there are two axioms shown in Figure~\ref{fig:axioms}. Their intuition is easily derived from the properties of equality and
the meaning of a sequent. Axiom $\AxiomRule$ for instance allows a branch of a formal proof to be closed when antecedent and succedent
of the current sequent contain a common formula. Such a sequent is clearly valid, given the conjunctive interpretation of the antecedent
and the disjunctive interpretation of the succedent. I.e.\ when every formula in the antecedent is satisfied, and one of them also occurs
in the succedent, then some formula of the succedent is also satisfied. The other axiom incorporates the reflexivity principle of equality:
when the goal is to show that some term equals itself or something else holds, then nothing more is to be proved in fact.

\begin{figure}[t]
\begin{center}
  \begin{prooftree}
  \AxiomC{}
  \LeftLabel{\AxiomRule}
  \UnaryInfC{$a = f(c), E(a,c), E(b,c) \derives a = f(c)$} 
  \LeftLabel{\SubstRightRule}
  \UnaryInfC{$a = f(c), b = f(c), E(a,c), E(b,c) \derives a = b$} 
  \UnaryInfC{(**)}  
  \end{prooftree}
  \vskip2mm
  
  \begin{prooftree}
  \AxiomC{}
  \LeftLabel{\AxiomRule}
  \UnaryInfC{$a = f(c), E(a,c), E(b,c) \derives E(b,c), a = b$} 
  \AxiomC{(**)}
  \LeftLabel{\ImpLeftRule}
  \BinaryInfC{$a = f(c), E(b,c) \imp b = f(c), E(a,c), E(b,c) \derives a = b$} 
  \UnaryInfC{(*)}
  \end{prooftree}
  \vskip2mm
  
  \begin{prooftree}
  \AxiomC{}
  \LeftLabel{\AxiomRule}
  \UnaryInfC{$E(b,c) \imp b = f(c), E(a,c), E(b,c) \derives E(a,c), a = b$}
  \AxiomC{(*)}
  \LeftLabel{\ImpLeftRule}
  \BinaryInfC{$E(a,c) \imp a = f(c), E(b,c) \imp b = f(c),  E(a,c), E(b,c) \derives a = b$}
  \LeftLabel{\ForallLeftRule}
  \UnaryInfC{$E(a,c) \imp a = f(c), \forall y.\; E(b,y) \imp b = f(y),  E(a,c), E(b,c) \derives a = b$}
  \LeftLabel{\ForallLeftRule}
  \UnaryInfC{$E(a,c) \imp a = f(c), \forall x\forall y.\; E(x,y) \imp x = f(y),  E(a,c), E(b,c) \derives a = b$}
  \LeftLabel{\ForallLeftRule}
  \UnaryInfC{$\forall y.\; E(a,y) \imp a = f(y), \forall x\forall y.\; E(x,y) \imp x = f(y),  E(a,c), E(b,c) \derives a = b$}
  \LeftLabel{\ForallLeftRule}
  \UnaryInfC{$\forall x\forall y.\; E(x,y) \imp x = f(y), \forall x\forall y.\; E(x,y) \imp x = f(y),  E(a,c), E(b,c) \derives a = b$}
  \LeftLabel{\ContrLeftRule}
  \UnaryInfC{$\forall x\forall y.\; E(x,y) \imp x = f(y),  E(a,c), E(b,c) \derives a = b$}
  \LeftLabel{\AndLeftRule}
  \UnaryInfC{$\forall x\forall y.\; E(x,y) \imp x = f(y),  E(a,c) \wedge E(b,c) \derives a = b$}
  \LeftLabel{\ImpRightRule}
  \UnaryInfC{$\forall x\forall y.\; E(x,y) \imp x = f(y) \derives E(a,c) \wedge E(b,c) \imp a = b$}
  \LeftLabel{\ForallRightRule}
  \UnaryInfC{$\forall x\forall y.\; E(x,y) \imp x = f(y) \derives \forall z.\; E(a,z) \wedge E(b,z) \imp a = b$}
  \LeftLabel{\ForallRightRule}
  \UnaryInfC{$\forall x\forall y.\; E(x,y) \imp x = f(y) \derives \forall y\forall z.\; E(a,z) \wedge E(y,z) \imp a = y$}
  \LeftLabel{\ForallRightRule}
  \UnaryInfC{$\forall x\forall y.\; E(x,y) \imp x = f(y) \derives \forall x\forall y\forall z.\; E(x,z) \wedge E(y,z) \imp x = y$}
  \end{prooftree}
\end{center}

\caption{A formal proof for the sequent given in Example~\ref{ex:exampleseq}.}
\label{fig:exampleproof}
\end{figure}

\begin{example}
\label{ex:exampleseq}
Figure~\ref{fig:exampleproof} depicts a formal proof of the valid sequent
\begin{displaymath}
\forall x\forall y.\; E(x,y) \imp x = f(y) \derives \forall x\forall y\forall z.\; E(x,z) \wedge E(y,z) \imp x = y\ .
\end{displaymath}
Intuitively, this sequent formalises the following assertion: if the inverse of the relation $E$ is included in the function $f$ then no 
element $z$ can have two different predecessors in the relation $E$. 

In order to formally prove this valid statement using the sequent calculus, can proceed as follows.\footnote{The proof shown in 
Figure~\ref{fig:exampleproof} is not the only but one of the shortest and simplest.} First we introduce fresh 
constants $a,b,c$ for the universally quantified variables in the antecedent using rule $\ForallRightRule$ thrice. This amounts to showing
that 
\begin{displaymath}
E(a,c) \wedge E(b,c) \imp a=b
\end{displaymath}
holds for arbitrary $a,b$ and $c$, provided that the antecedent -- untouched up until then -- holds, too. It allows us to use an
implicit property of functions, namely that they map values to values uniquely, for the relation $E$. However, we need to use the
assumption twice, once for $a$ and $c$ but also for $b$ and $c$. This is why the proof uses rule $\ContrLeftRule$ and duplicates
this assumption before we instantiate the universally quantified formulas in the antecedent with these two pairs of constants (which
are now available as ground terms). This is an insight one needs to have at this point for otherwise one will not be able to close
the proof. Note that the application of rule $\ContrLeftRule$ is not driven by any syntactical need like a top-level operator in 
the sequent. 

The rest of the proof -- which is broken into three parts in order to fit the page width here -- uses a bit of simple Boolean reasoning
eliminating the implication connectors, followed by very simple equational reasoning basically implementing a standard pattern to show
that equality is transitive using rule $\SubstRightRule$.
\end{example}

\subsection{The Sequent Calculus as a Formal Proof System}

We briefly discuss two aspects of the sequent calculus. The first one is purely mathematical at first sight and can be phrased as 
follows.

\begin{proposition}[\cite{Gentzen35a, eft_logic_book84}]
The Sequent Calculus is sound and complete with respect to validity of sequents in First-Order Logic with Equality.
\end{proposition}

Soundness means that any sequent which can be derived, i.e.\ for which there is a proof, is indeed valid. In other words, one cannot
prove false statements using the sequent calculus. It can be shown by induction on the height of a proof tree, making use of the fact
that all the rules are valid in the sense that a conclusion is valid if all its premisses are valid. In the special case of axioms this
means that conclusions are valid straight away. Completeness means that any valid sequent is provable. This is more difficult to show, 
c.f.\ \cite{eft_logic_book84}. 

These meta-logical principles about the sequent calculus have some effect on didactical questions. Completeness of the calculus guarantees
that a proof exists for a sequent which may -- by informal reasoning -- be seen as valid. So any student's difficulty in being able to
construct a proof is down to the student's abilities and experience with this calculus. It does not introduce an additional level of
difficulty by requiring a student to understand for which valid sequents formal proofs could be constructed and for which it simply is
impossible. Soundness is of course at least equally important as working with an unsound proof calculus would severely impede students'
abilities to distinguish valid from invalid statements and therefore also to learn how to read off the intuitive meaning of formulas.

The second aspect worth mentioning here concerns the format of proofs in the sequent calculus as opposed to other proof calculi for
First-Order Logic with Equality. The application of a rule in the sequent calculus is purely local, as opposed to natural deduction
for instance where rule application can span over the entire history of formal statements from the beginning of the proof attempt. The
rules of the sequent calculus carry all assumptions and proof goals through to the premisses such that a decision on which rule to apply
next can (at least in principle) be taken purely based on a current sequent alone, disregarding all other sequents of a partially 
constructed formal proof. 


\section{Learning the Sequent Calculus}

\subsection{Syntactic Rule Manipulation vs. Semantic Understanding}
\label{didactics}

A standard exercise in sequent calculus asks for a proof of a given sequent. We reconsider the valid sequent $\mathcal{S}_0$ presented in
Example~\ref{ex:exampleseq} above:  
\begin{displaymath}
	\forall x\forall y.\; E(x,y) \imp x = f(y) ~ \yields ~ \forall x\forall y\forall z.\; E(x,z) \wedge E(y,z) \imp x = y\ .
\end{displaymath}
The students' difficulties when trying to find a formal proof for such sequents, as the one given in Figure~\ref{fig:exampleproof} for 
instance, as well as mistakes frequently made when trying to construct such a proof on paper can generally be put into two categories --
the syntactic and semantic ones mentioned in the introduction. 
\medskip

(1) The first one is about \emph{constructing a correct proof}: many students are not able to handle formalisms well; often they can 
barely parse sequents and apply rules correctly. In this example, one has to introduce new names for the universally quantified 
variables $x,y,z$ in this order -- say $a,b,c$ -- and then decompose the Boolean operators on the right side, yielding $\mathcal{S}_1 :=$
\begin{displaymath}
	\forall x\forall y.\; E(x,y) \imp x = f(y), E(a,c), E(b,c) \derives a = b\ .
\end{displaymath}
Typical mistakes at this syntactic level are concerned with wrong rule applications and include 
\begin{itemize}
	\item \emph{confusing} rules, for instance applying the rule for conjunctions to a disjunction;
	\item \emph{misplacing} rules, usually by applying a rule to a genuine subformula rather than a formula in the sequent; in other words not
	understanding the structure of a sequent;
	\item \emph{wrong first-order instantiations}, for instance not choosing a fresh Skolem constant when needed;
	\item \emph{wrong rule instantiations}, for instance by adding the symbols $\Gamma$ and $\Delta$ to the sequent at hand; 
\end{itemize}
and so on \cite{EHL:TTL15}.
\medskip

(2) The second category concerns the semantical understanding of a proof, which we here address as the ability of \emph{finding the right proof}.
We exemplify the difficulties in finding a proof for sequent $\mathcal{S}_0$ from above.
As mentioned above, one has to start by applying a ``purely syntactic'' strategy, which eventually yields the sequent $\mathcal{S}_1:=$
\begin{displaymath}
\forall x\forall y.\; E(x,y) \imp x = f(y),\; E(a,c),\; E(b,c) ~ \yields ~  a = b\ .
\end{displaymath}
At this point, the next step is not obvious. Clearly, based on results on semi-decidability and completeness of the sequent calculus \cite{eft_logic_book84} there is a strategy that will always find a proof if one exists: it consists of applying every possible rule instance at some point. This is not a good strategy for students in an exam or homework-assignment, though, as it simply takes far too long to find a proof, and the resulting proofs are too large to be overseen by the human mind. Thus, this generic strategy is hardly useful to yield semantic understanding, let alone solve homework or exam exercises.

A closer look at the sequent can give the right intuition needed to find a short proof. In analogy to the explanations in Section~\ref{subsec:formal_proofs} the sequent $\mathcal{S}_1$ can be rephrased as follows: \textit{Given a graph $G=(E,V)$, if the inverse of the edge-relation $E$ is functional and there are edges from $a$ to $c$ and $b$ to $c$, then $a$ has to equal $b$}.

Now, a na\"{\i}ve approach to prove sequent $\mathcal{S}_1$ is to continue with obvious rule applications like instantiating the universal quantifiers in the premiss, e.g.\ resulting in the sequent $\mathcal{S}_2:=$
\begin{displaymath}
E(b,c) \imp b = f(c),\; E(a,c),\; E(b,c) ~ \yields ~  a = b\ .
\end{displaymath}

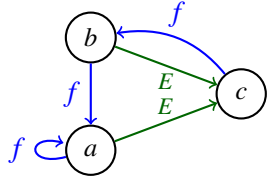
\begin{figure}[t]
	\begin{center}
		\begin{tikzpicture}[thick]
		\draw[draw=none,use as bounding box] (-0.5,-0.5) rectangle (2.5,2.2);
		\node[draw,circle,minimum size =1.7em] (a) at (0,0) {\small $a$};	\node[draw,circle,minimum size =1.7em] (b) at (0,1.5) {\small $b$};
		\node[draw,circle,minimum size =1.7em] (c) at (2,0.75) {\small $c$};
		
		\path[->,draw,black!60!green] (a) edge node[above=-1pt] {\footnotesize $E$} (c)
		(b) edge node[below=-1pt] {\footnotesize $E$} (c);
		
		\path[->,blue] (c) edge[bend right] node[above] {$f$} (b)
		               (b) edge             node[left]  {$f$} (a)
		               (a) edge[loop left]  node[left]  {$f$} ();
		\end{tikzpicture}
	\end{center}
	\caption{Countermodel for $\mathcal{S}_2$.}
	\label{countermodel}
\end{figure}

This sequent has a simple countermodel, shown as a directed graph in Figure~\ref{countermodel} with edges depicting the relation $E$ in green
and the function $f$ in blue. 

Thus, the valid sequent $\mathcal{S}_1$ has turned into the invalid sequent $\mathcal{S}_2$ by an application
of a particular proof rule, and it should be clear that such a rule application is to be deemed as a bad step in trying to construct a
formal proof for a valid sequent.

Actually, the existence of a countermodel to $\mathcal{S}_2$ provides the right intuition to prove sequent $\mathcal{S}_1$. We have to 
use the premiss $\forall x\forall y.\; E(x,y) \imp x = f(y)$ on all edges of the given graph. The advisable next step is therefore to 
double the formula $\forall x\forall y.\; E(x,y) \imp x = f(y)$ using rule \ContrLeftRule and then apply it also to the edge from $a$ to $c$. The resulting sequent is $\mathcal{S}_3 :=$
\begin{displaymath}
	E(a,c) \imp a = f(c),\;E(b,c) \imp b = f(c),\; E(a,c),\; E(b,c) \derives a = b\ ,
\end{displaymath}
which is easily proved by decomposing Boolean operators and substituting equal terms using the corresponding rules.

\subsection{Tool-Supported Learning}

There is a clear dependency between the two challenges mentioned in Section~\ref{didactics}: the syntactic ones described in (1) need to 
be met before the semantic ones in (2); it is impossible to find a proof unless one is able to construct correct proofs at all. The latter 
is clearly a very difficult task for students who already struggle with uncertainties like ``am I allowed to apply this rule here?'', 
``was the application correct?'', ``should I introduce a new name or instantiate with an already existing term?'', etc.

This gap between syntactical and semantical understanding has been addressed in many fields like teaching programming or mathematics 
(e.g.\ in \cite{DBLP:journals/ijpp/ShneidermanM79}). The phenomenon is accurately described in \cite{DBLP:conf/ticttl/GasquetSS11a} as 
the ability to ``write rather rigorously a simple C program'', while they cannot ``rigorously write down a mathematical proof of the 
kind needed in graph theory, formal logic, [...]''  There is a hidden hint in this observation on how to tackle the problem of teaching 
proof calculi. Students seem to easily understand syntactic principles as long as there is a mechanism -- like a compiler -- which allows 
them to learn the formalism in a trial-and-error way.

This is where the Sequent Calculus Trainer (SCT) comes into play. It is supposed to aid the students in facing the aforementioned 
challenges of constructing correct proofs and finding the right proof. SCT therefore adheres to two design principles: it is 
an easy-to-use and simple assistant for building proof trees in the sequent calculus, while also providing compiler-like feedback on
syntactical rule applications which is known for its benefits in tutorial teaching environments \cite{Anderson1995}. 
Moreover, it offers an interactive proof mode, meant to guide the students' focus on the underlying semantics of a sequent. 

\subsection{Related Tools}\label{subsec:related_tools} 
There are other high quality interactive proof systems that can be used to train the construction of proofs in the sequent calculus. Here we only mention three, which draw our attention through their accessibility and usability.

The tool that meets the prerequisites laid out here best is LOGITEXT\footnote{\url{http://logitext.mit.edu}}; others worth mentioning are JAPE \cite{DBLP:journals/cj/BornatS99} and KEY\footnote{\url{http://www.key-project.org}}. One major drawback of the two mentioned first is that they don't treat first-order logic with equality out of the box, though, JAPE offers the possibility to add rules for treating equality to its sequent calculus. Although, the KEY-System is meant as a verification tool for JAVA-Programs it can be used as an interactive prover of sequents in first-order logic.

However, none of those existing tools is general enough to serve the subtle didactical purposes described above -- to act as a tool that is meant to trigger semantical understanding in learning how to proof in an error-guided fashion.
Furthermore, in a setting where we also measure success through the understanding of syntactic concepts, it is essential that the proof calculus used in classroom must
be the same as that used by a supporting tool. Thus, one should not underestimate the effort that would be needed in order to extend or amend an existing software tool created by others. Hence, having many tools with slightly differing features
in this area should be considered advantageous.


\section{The Sequent Calculus Trainer}\label{seq:sct}

\begin{figure}
	\begin{center}
		\includegraphics[width=\textwidth]{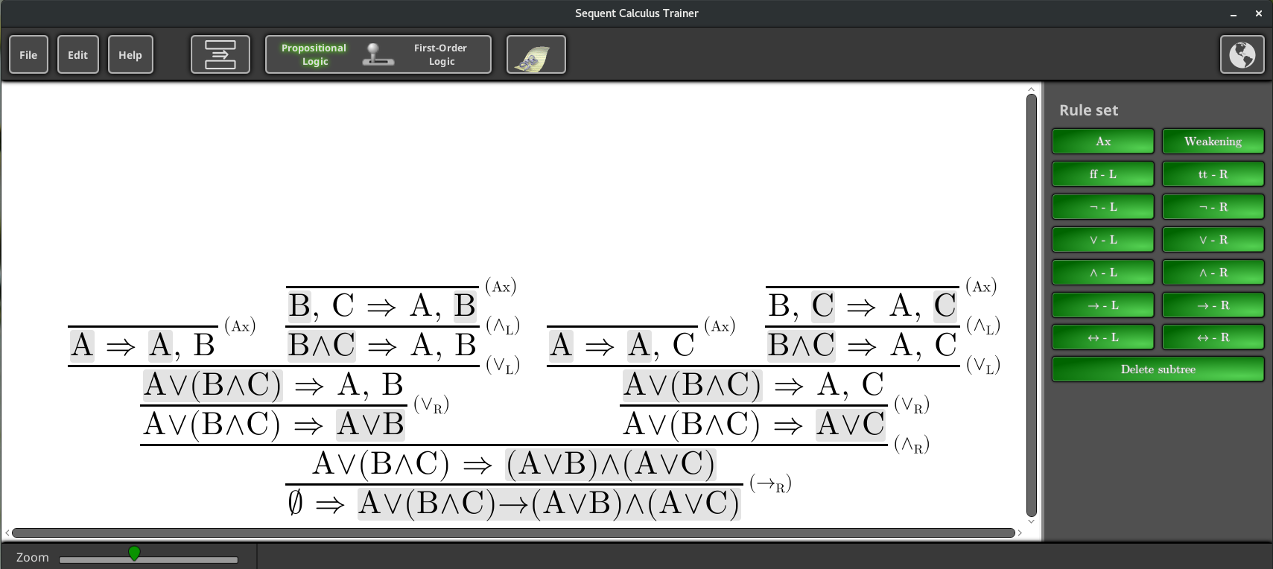}
		\caption{The user-interface.}\label{ui}
	\end{center}
\end{figure}

We provide the Sequent Calculus Trainer (SCT) as an open source application under the BSD-3 license. The source code as well as the binaries are publicly
available.\footnote{\url{http://www.uni-kassel.de/eecs/fachgebiete/fmv/projects/sequent-calculus-trainer.html}} Figure~\ref{ui} shows the graphical user interface which is kept fairly
simple.

\begin{figure}
	\begin{center}
		\includegraphics[width=\textwidth]{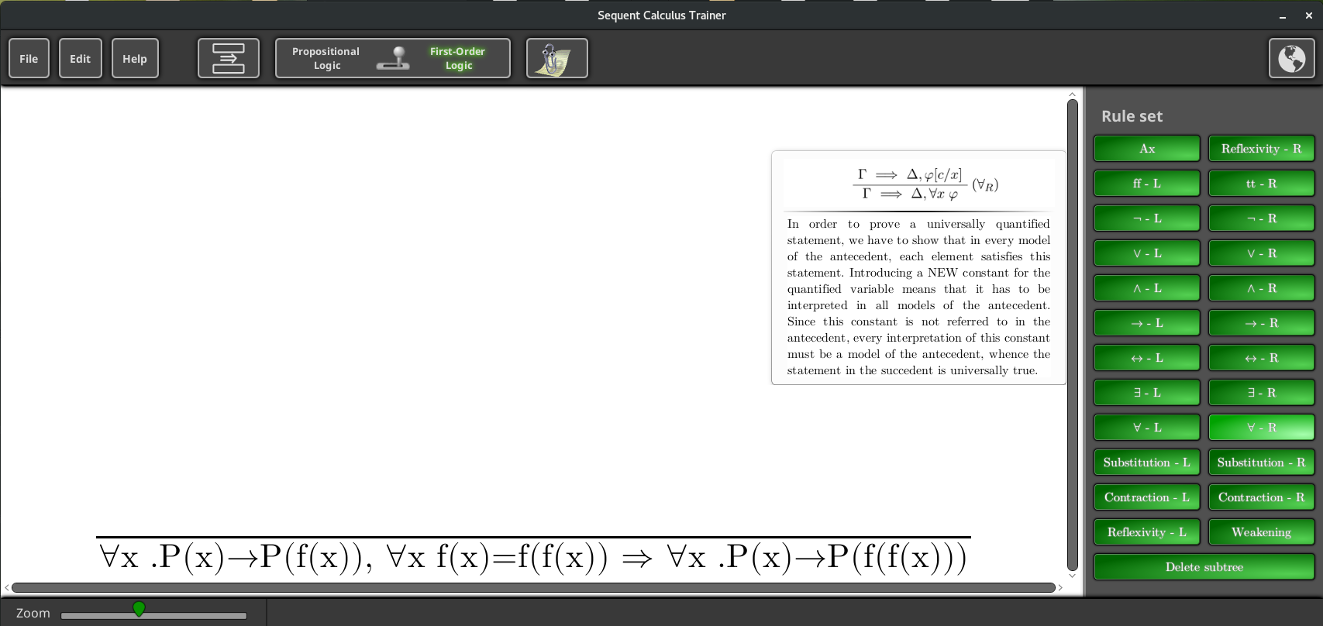}
		
		\vspace*{0.8em}
		\includegraphics[width=\textwidth]{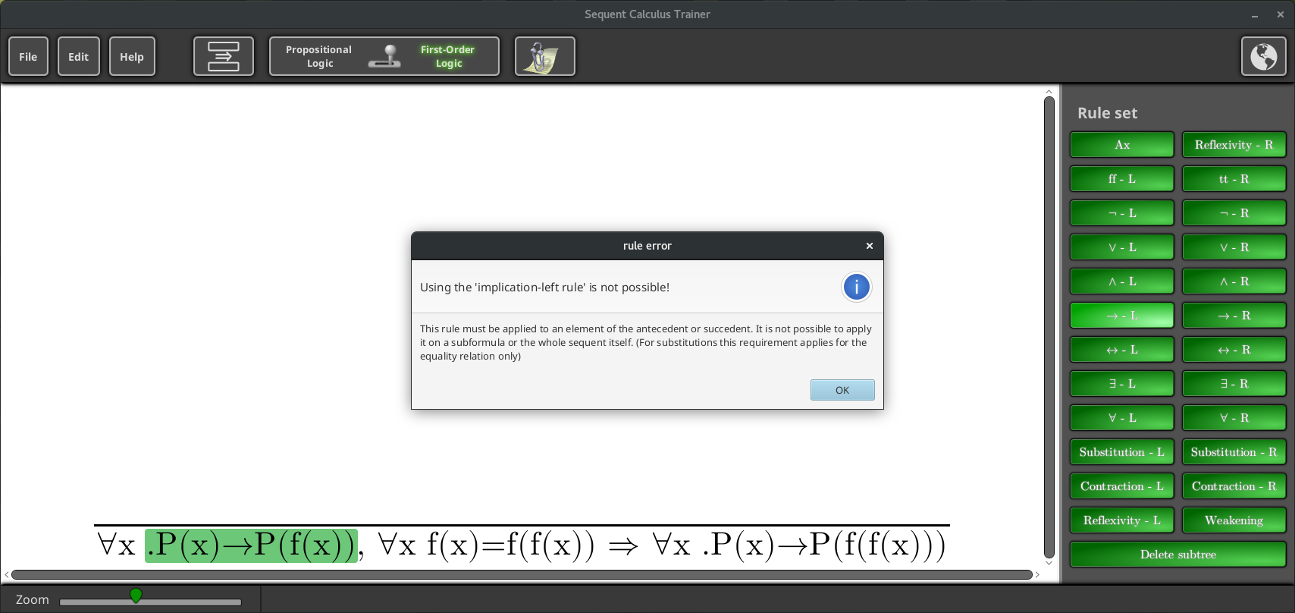}
		\caption{The feedback system.}\label{feedback}
	\end{center}
\end{figure}

SCT comes with two main views, one for propositional logic and one for first-order logic. They both differ only in the number of
applicable rules shown on the right side of the windows and in the treatment of atomic propositions which are interpreted as 0-ary 
predicates in the first-order logic view. Sequents can be input either through a text file or via a text field where the syntax 
specification for the input is given, too. Furthermore, it is possible to save and load proof trees in an internal format as well as
export them in PNG format.

In the following we introduce the key features of the Sequent Calculus Trainer distinguished by their purpose of helping students develop syntactical understanding and semantical understanding for the task of finding formal proofs in the sequent calculus.

\subsection{Support for Constructing Syntactically Correct Proofs}
The Sequent Calculus Trainer can be used as a simple assistant for constructing proof trees, without providing any hints on which rules 
to apply. In the following we will briefly introduce the two key features of this mode.

\begin{figure}
	\begin{center}
		\includegraphics[width=\textwidth]{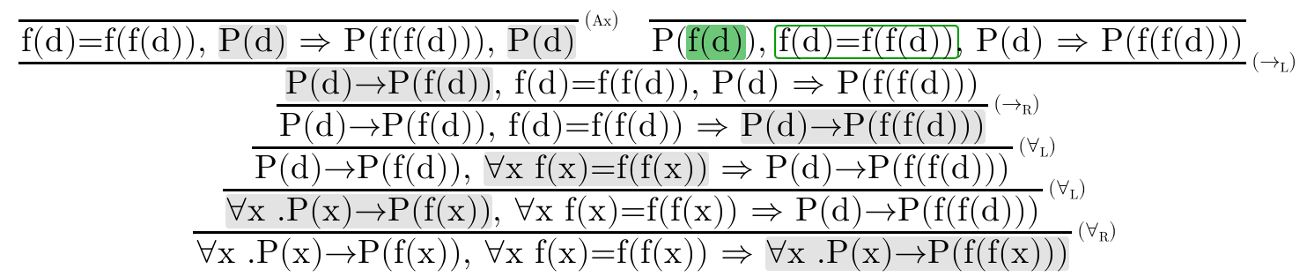}
		
		\vspace*{0.8em} 
		\includegraphics[width=\textwidth]{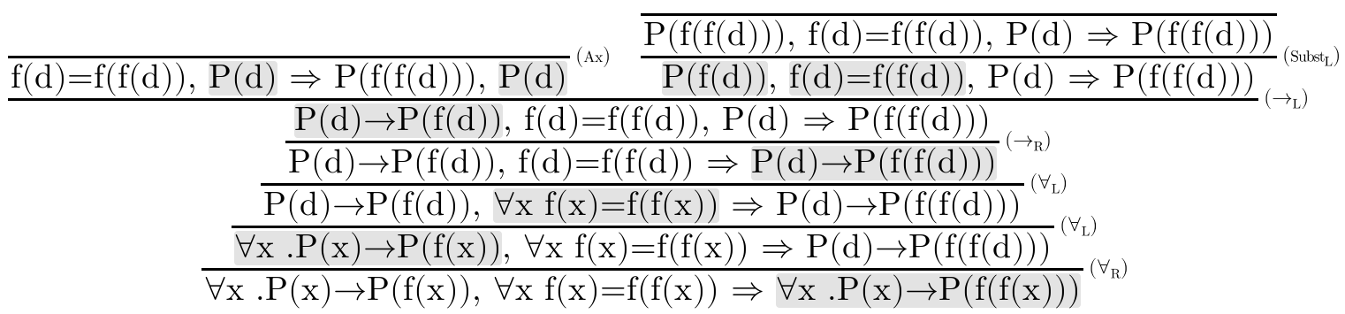}
		\caption{Applying the substitution rule in a proof: before (top) and after (below).}\label{equality}
	\end{center}
\end{figure}

\begin{itemize}
	\item Nearly every user action leads to a response by the program.  Figure~\ref{feedback} exemplarily illustrates such on-screen messages. Each rule
	button is equipped with a short message which occurs on mouse-over. These messages usually contain the formal definition of a rule as well as
	some appropriate high-level explanations of the rule's meaning and, if suitable, why it is a valid logical principle. 
	
	If the user has chosen a rule and tries to apply it to a formula by selecting a logical operator the formula represented by this operator ``responds'' 
	by telling the user whether or not the rule is applicable there. This happens in two ways: the part of the formula that is in scope of the selected 
	operator or symbol is highlighted. This helps to understand precedence rules and the structure of sequents and formulas.  When a wrong operator or 
	symbol is chosen the user is provided with an error message which includes a hint on the mistake. For instance, the reason why a current leaf in 
	the proof tree is an axiom has to be identified via clicking on the part of the formula that causes the application of an axiom rule.
	
	\item The second notable feature is the handling of sequents that include equalities. Figure~\ref{equality} shows how the substitution rule works. 
	After the rule for substitution on the left-hand or right-hand side of the sequent has been chosen the program expects an atomic formula with an 
	equality predicate to be selected. In the last step, the term that should be substituted needs to be clicked on.
\end{itemize}
A final point worth mentioning is that the user is able to undo all steps in the proof up to a certain sequent at any time by just applying a different 
rule to that particular sequent. 

\subsection{Support for Finding the Right Proof}

\begin{figure}
	\begin{center}
		\includegraphics[width=0.9\textwidth]{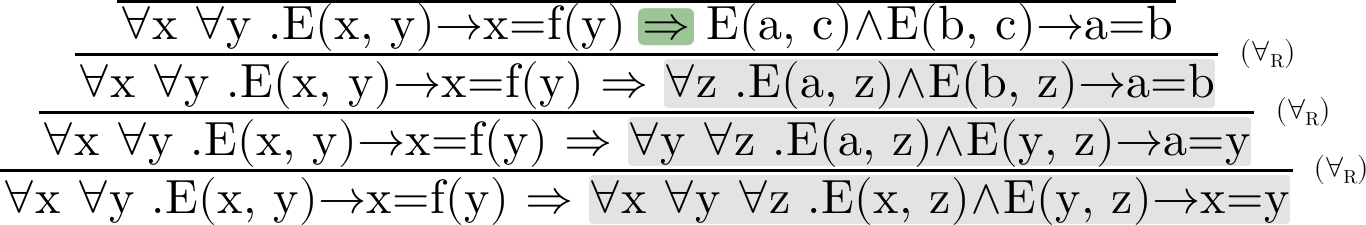}
		\caption{The beginning of a proof for the sequent $\mathcal{S}_1$.}\label{s1}
	\end{center}
\end{figure}

The interactive proof mode uses a traffic-light-like system to provide students with feedback to indicate whether the current sequent is valid (green), invalid (red), or of unknown status (yellow). Fig.~\ref{s1} shows the situation for the sequent $\mathcal{S}_0$ from 
Section~\ref{didactics}.

\begin{figure}
	\begin{center}
		\includegraphics[width=0.9\textwidth]{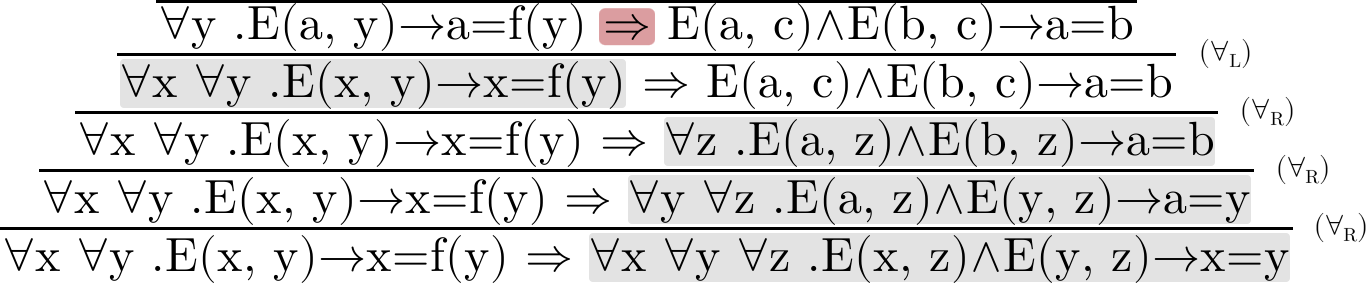}
		\caption{Creating an invalid sequent by unwise instantiation.}\label{inval}
	\end{center}
\end{figure}

As soon as one quantifier in the premiss is resolved without doubling the premiss first the sequent becomes unprovable. This is announced to the user by turning the indication light red, as shown in Fig.~\ref{inval}.

So the interaction between SCT and a user is thus governed by a very simple protocol: SCT immediately reports to the user what the effect of the
last rule application was in terms of chances to finish the formal proof in this way. The information transported through this traffic-light
system is the following: as soon as a valid sequent turns into an invalid one there is no hope of completing the proof in this way; so the 
student has to rethink the last step. In general, he/she cannot simply try out all possible rule applications at this point because
there is no a priori bound on 
\begin{itemize}
\item the number of duplicates one may need using rules \ContrLeftRule\ or \ContrRightRule, and on
\item the number of ground terms that suffice for instantiations using rules \ForallLeftRule\ and \ExistsRightRule.
\end{itemize} 
Furthermore, as it is clear that no syntactic mistake has been made at this point the best strategy for completing a proof is to understand
what went wrong in the rule application that turned a green light red. This initiates a process of deeper semantical understanding, for instance
by forcing the student to consider a counter model for the invalid sequent or understand the intuitive reason for the validity of the
implication present in the previous sequent. 

For valid sequents the Sequent Calculus Trainer also possesses a mode in which information on which rule to apply next is provided to the
students.

\section{Behind the Scenes}

The Sequent Calculus Trainer is a Java implementation; this guarantees accessibility for a wide range users. It uses the GUI framework 
JavaFX\footnote{\url{https://docs.oracle.com/javafx/2/overview/jfxpub-overview.htm}} which is integrated in the Java Standard Library since the emerging of Oracle's Java 
8. 

In the following, we will describe the main ideas behind automated reasoning algorithm used to determine the display colour for a sequent
in SCT's interactive helper mode. Remember that validity for sequents of First-Order Logic is undecidable (c.f. \cite{eft_logic_book84}). Hence, there is no
obvious algorithm -- let alone any -- for determining validity and turning its answer into either of the colours red or green. On the other
hand, the validity problem is semi-decidable (c.f. \cite{eft_logic_book84}) which is typically shown in a way that boils down to arguing that the set of 
finite proof trees is recursively enumerable. However, a brute-force method of enumerating all proof trees, suitably combined with a time-out, 
hardly yields a practical procedure for displaying the validity status of a sequent through the colours red, green and yellow. The didactical
purpose that SCT serves requires the validity test to be efficient and yield the colour yellow in as few cases as possible, for otherwise
SCT faces the danger of not being accepted by students as a helpful tool. 

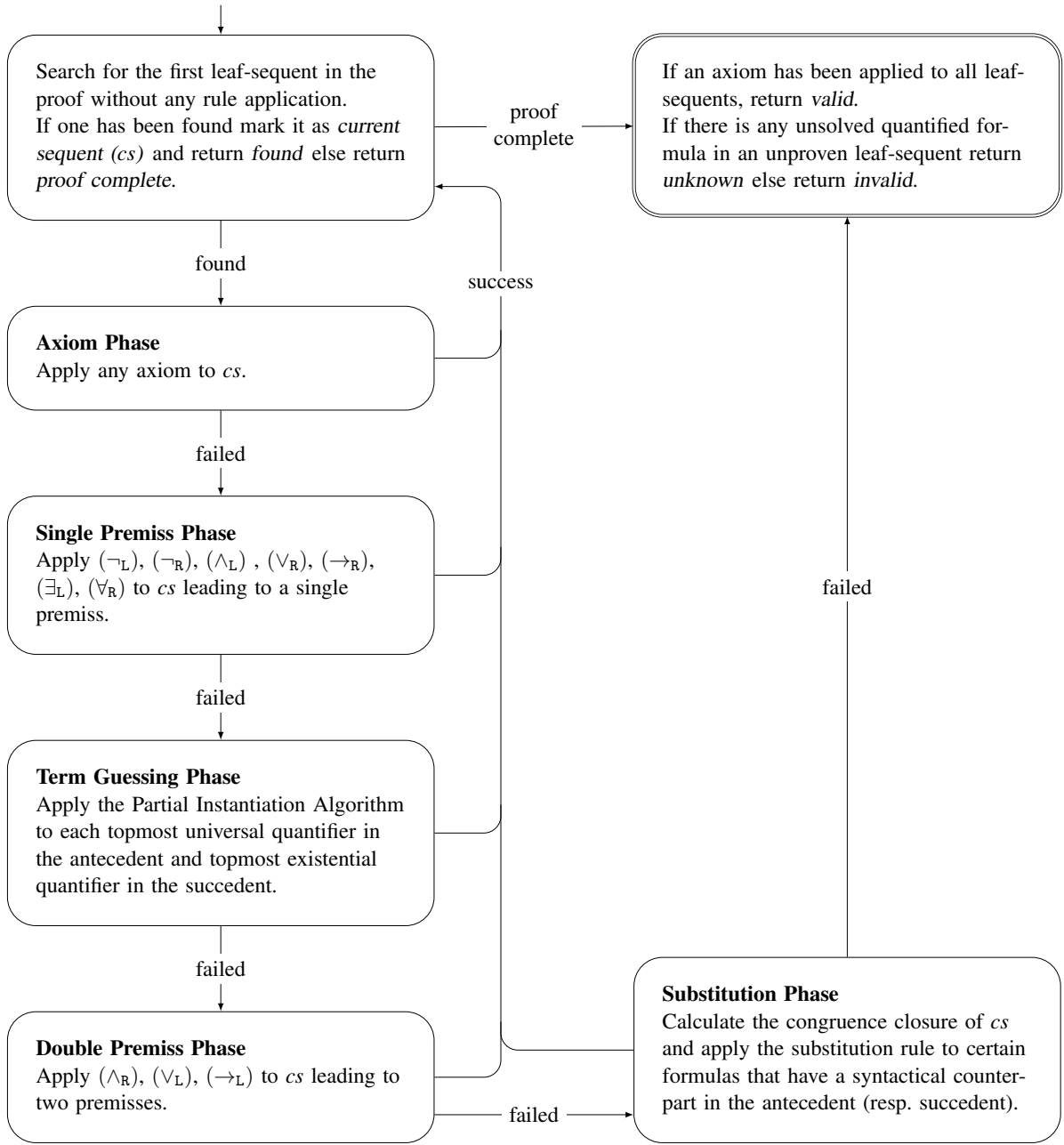
\begin{figure}
	\begin{center}
	\scalebox{0.85}{
	\begin{tikzpicture}
	
	\tikzset{>=latex}
	
	\tikzset{
		textbox/.style={draw=black, fill=white, rectangle, rounded corners=5mm, text width=65mm, inner sep=5mm},
		doubletextbox/.style={draw=black, double, double distance=1pt, fill=white, rectangle, rounded corners=5mm, text width=65mm, inner sep=5mm}
	}
	
	\edef\boxvertdist{15mm}
	\edef\boxhordist{35mm}
	\edef\loneboxshift{6mm}
	\edef\lfiveboxshift{5mm}
	\edef\harpoonlength{8pt}

	\node[textbox, initial above, initial text=] (l1) {\parbox{\linewidth}{Search for the first leaf-sequent in the proof without any rule application.\\
			If one has been found mark it as \textsl{current sequent ($cs$)} and return \textsl{found} else return \textsl{proof complete}.}};
	
	\node[textbox] (l2) [below=\boxvertdist of l1] {\parbox{\linewidth}{\textbf{Axiom Phase}\\
			Apply any axiom to $cs$.}};
	
	\node[textbox] (l3) [below=\boxvertdist of l2] {\parbox{\linewidth}{\textbf{Single Premiss Phase}\\
			Apply $\NegLeftRule$, $\NegRightRule$, $\AndLeftRule$ , $\OrRightRule$, $\ImpRightRule$, $\ExistsLeftRule$, $\ForallRightRule$ to $cs$ leading to a single premiss.}};
	
	\node[textbox] (l4) [below=\boxvertdist of l3] {\parbox{\linewidth}{\textbf{Term Guessing Phase}\\
			Apply the Partial Instantiation Algorithm to each topmost universal quantifier in the antecedent and topmost existential quantifier in the succedent.}};
	
	\node[textbox] (l5) [below=\boxvertdist of l4] {\parbox{\linewidth}{\textbf{Double Premiss Phase}\\
			Apply $\AndRightRule$, $\OrLeftRule$, $\ImpLeftRule$ to $cs$ leading to two premisses.}};

	\node[doubletextbox] (r1) [right=\boxhordist of l1.north east, anchor=north west] {\parbox{\linewidth}{If an axiom has been applied to all leaf-sequents, return \textsl{valid}.\\
			If there is any unsolved quantified formula in an unproven leaf-sequent return \textsl{unknown} else return \textsl{invalid}.}};
	
	\node[textbox] (r2) [right=\boxhordist of l5.south east, anchor=south west] {\parbox{\linewidth}{\textbf{Substitution Phase}\\
			Calculate the congruence closure of $cs$ and apply the substitution rule to certain formulas that have a syntactical counterpart in the antecedent (resp.\ succedent).
	}};

	\draw[->] (l1) -- node[fill=white, align=center] {proof \\ complete} (r1);
	
	\coordinate (l1southeast) at ([xshift=\boxhordist / 3, yshift=-3mm]l1.south east);
	\coordinate (l2east) at ([xshift=\boxhordist / 3, yshift=5mm]l2.east);
	\coordinate (l3east) at ([xshift=\boxhordist / 3, yshift=5mm]l3.east);
	\coordinate (l4east) at ([xshift=\boxhordist / 3, yshift=5mm]l4.east);
	\coordinate (r2west) at ([xshift=\boxhordist]l4.east);
	\coordinate (l5east) at ([xshift=\boxhordist / 3, yshift=5mm]l5.east);
	\coordinate (l6east) at ([xshift=-(\boxhordist- \boxhordist / 3), yshift=5mm]r2.west);
	
	\draw[<-, rounded corners=3mm] ([yshift=\loneboxshift]l1.south east) -| (l1southeast);
	\draw (l1southeast) -- node[fill=white, align=center] {success} (l2east);
	\draw[rounded corners=3mm] (l2.east) -| (l2east);
	\draw (l2east) -- (l3east);
	\draw[rounded corners=3mm] (l3.east) -| (l3east);
	\draw (l3east) -- (l4east);
	\draw[rounded corners=3mm] (l4.east) -| (l4east);
	\draw[rounded corners=3mm] (r2.west) -| (l6east);
	\draw (l4east) -- (l5east);
	\draw[rounded corners=3mm] (l5) -| (l5east);
	
	\draw[->] ([yshift=\lfiveboxshift]l5.south east) -- node[fill=white, align=center] {failed} ([yshift=\lfiveboxshift]r2.south west);
	
	\draw[->] (l1) -- node[fill=white, align=center] {found} (l2);
	\draw[->] (l2) -- node[fill=white, align=center] {failed} (l3);
	\draw[->] (l3) -- node[fill=white, align=center] {failed} (l4);
	\draw[->] (l4) -- node[fill=white, align=center] {failed} (l5);
	
	\draw[->] (r2) -- node[fill=white, align=center] {failed} (r1);
	\end{tikzpicture}}
	\end{center}
	\caption{Partial Instantiation Algorithm.}
	\label{fig:algo}
\end{figure}

The main techniques used in the implementation of the validity test concern an efficient instantiation of quantified variables
followed by a decision procedure for the validity of quantifier-free first-order formulas with equality. The latter problem is well-known
to be decidable, for instance using a result on the decidability of the congruence closure of equality terms (e.g.\ in 
\cite{Baader:1998:TR:280474}). Hence, we focus on the instantiation problem here. Figure~\ref{fig:algo} shows the algorithm's general structure. 
A comprehensive reflection of all techniques used can be found in \cite{Ehle17}.

The principle idea underlying the validity test uses the following connection between invalidity of sequents and satisfiability of formulas:
\begin{displaymath}
\mathcal{S} = \Gamma \Longrightarrow \Delta\ \text{ invalid} \quad \Longleftrightarrow \quad 
\phi_{\mathcal{S}} := \bigwedge_{\psi\in\Gamma} \psi \land \bigwedge_{\psi'\in\Delta} \lnot \psi'\ \text{ is satisfiable.}
\end{displaymath}
Hence, it is principally possible to use a satisfiability solver for First-Order Logic for this task. This solver needs to be able to handle
quantifiers and uninterpreted function symbols. The former is supported by some, the latter by most modern SMT solvers.\footnote{\ldots even
though the name \emph{Satisfiability Modulo Theories} \cite{series/faia/BarrettSST09} indicates that they were predominantly designed to 
solve satisfiability problems for formulas over interpreted function symbols like addition on numbers for instance.} We make use of Microsoft
Research's high performing solver Z3 \cite{DeMoura:2008:ZES:1792734.1792766} in version 4.5.0.

However, there is an apparent mismatch in using this reduction. A satisfiability solver like SMT either reports unsatisfiability, a time-out
or satisfiability. In the latter case, Z3 produces a model for the formula; in the former case no witness for unsatisfiability is given, 
though. This does not go well with the requirements laid out for SCT's interactive helping mode (and also has to do with results on the
semi-decidability of validity, resp.\ of satisfiability over finite models for FOL): it is the case of validity in which a witness would
be needed in order to turn this into a hint at which rule to apply next and how to do so. The witness obtained from the SMT solver in case
of satisfiability can of course only be used for a hint about invalidity; in fact, such a witness reported by the SMT solver is simply a 
counter model for the sequent.

In order to facilitate a hint at which rule to apply to a valid sequent next we use the following trick. A quantified variable (universally
in the antecedent or existentially in the succedent) is instantiated with a term that is not a ground term. We call such an instantiation
\emph{partial}. Note that any non-ground term can be seen as a partially constructed ground term; those parts which have not been constructed
yet are being abbreviated by a variable. By successively refining partial instantiations it is possible to approximate ground terms as the
following simple example shows. The procedure building on this principle in order to test sequents for validity -- the Partial Instantiation
Algorithm -- is presented in Figure~\ref{fig:algo}.

\begin{example}
Consider the valid sequent
\begin{displaymath}
\forall y. P(g(e,f(y))), Q(f(c)), R(h(d,c),e) \derives \exists x\; P(x)\ .
\end{displaymath}
The last two formulas of the antecedent clearly do not contribute to the validity of this sequent; they are added just to blow up the 
underlying signature which helps to exemplify the problem with an approach based on a brute-force enumeration of all ground terms. Note
that this sequent can be proved using rule $\ExistsRightRule$ with the instantiation $x \mapsto g(e,f(d))$ for example, followed by
using rule $\ForallLeftRule$ with the instantiation $y \mapsto d$. In fact, any instantiation $x \mapsto g(e,f(t))$ and $y \mapsto t$
is sufficient for any ground term $t$.

Consider the following enumeration of all ground terms over the signature 
$\langle f^{(1)}, g^{(2)}, h^{(2)}, c^{(0)}, d^{(0)},$ $e^{(0)} \rangle$ which is obtained by using a standard diagonal search through the
multi-dimensional space of all ground terms:
\begin{displaymath}
c, d, e, f(c), f(d), g(c,c), f(e), g(d,c), h(c), g(e,c), h(d), f(f(c)), h(e), \ldots
\end{displaymath}
The first term that can be used to instantiate $x$ is $g(e,f(c))$, and it occurs in position 1049 of this enumeration. Even though Z3 
can speedily detect unsatisfiability of the instantiated formula 
\begin{equation}
\label{eq:satformula}
\forall y. P(g(e,f(y))) \wedge  Q(f(c)) \wedge R(h(d,c),e) \wedge \neg P(t)
\end{equation}
for any ground term $t$ that is not of the form $g(e,f(\ldots))$ this is not in any way efficient since
the overhead for building formulas and calling Z3 costs too much time.
\end{example}

Partial instantiation tries to find the correct instantiation value for variables like $x$ in the previous example by building such terms
in a goal-directed fashion. Instead of going through all ground terms according to some fixed enumeration we consider the variable at hand
as an already partially instantiated term. Then we take its subterm which is highest in the syntax tree and consists of a variable (i.e.\ 
at the beginning the term itself) and replace it successively by terms of the form $f(x_1,\ldots,x_n)$ for some $n$-ary function symbol
$f$ and variables $x_1,\ldots,x_n$. This way, the correct term for the instantiation is found by successive refinement. 

\begin{example} (Continued)
Using partial instantiation on the example formula above the SMT solver Z3 is successively asked to check satisfiability of the formula
(\ref{eq:satformula}) for the following partially instantiated values of $t$:
\begin{displaymath}
z, f(z), g(z, z'), h(z), f(f(z)), f(g(z,z')), f(h(z)), g(f(z),z'), g(g(z,z''),z'), g(h(z),z'), g(z,f(z')), \ldots
\end{displaymath}
where $z,z',z''$ are variables. Note how they are being obtained: starting with the maximally partially instantiated term $z$, we
replace its top-most variable by $f(z)$, $g(z,z')$ and $h(z)$ since $f,g,h$ are the only three function symbols in the underlying
signature. The next three terms are obtained by replacing $z$ in $f(z)$ with these three again. Then we get 9 more terms obtained
by refining $g(z,z')$, replacing each $z$ and $z'$ with either of these three, and so on.

As one can see, the (partially instantiated) term $g(z,f(z'))$ that leads to unsatisfiability of (\ref{eq:satformula})
occurs in position 11 of this list. It is then not hard to see that a proper ground term can be obtained from a partially instantiated
term that causes unsatisfiability, by replacing its variables with some combination of constants. Taking this into account, the first 
term that can be used to prove validity using \ExistsRightRule\ is found after trying out 74 instantiations only. Hence, only a fraction
of the number of calls to Z3 is needed -- at least in this case -- in comparison to the brute-force method of enumerating all ground terms.  
\end{example}
 
Once Z3 reports unsatisfiability in this way, a term has been constructed by successive refinements, and this term can be used to give the
student a hint as to how to use rule \ExistsRightRule  (and \ForallLeftRule in general, too).

Obviously, the chance of finding a proof depends on the success rate of the used SMT solver. Tests show that the running time of Z3 increases
especially with the number of different function symbols and the alternation of these symbols in terms. The smallest example we have found 
that makes Z3 fail to compute an answer is the formula
\newcommand{\forallx}[1]{\ensuremath{\forall{#1}\;}}
$\forallx{x}  P(f(g(f(x)))) \land \lnot P(f(g(h(c))))$.
Such cases are handled in the implementation by a fixed timeout; the GUI's feedback to the user is the yellow marker then indicating an 
unknown validity statement of the sequent at hand.


\section{Conclusion and Future Work}

In \cite{EHL:TTL15} we presented some statistical findings hinting that already the first version of the Sequent Calculus Trainer led to a significant improvement of student's learning outcomes in constructing correct proofs. Although the extension introduced here has not yet been tested in the setting of a major course on formal logic, we strongly believe that the current enhancement will initiate the desired thought processes which are needed for understanding the semantic structure behind a proof.

The description of the Partial Instantiation Algorithm gives rise to a more or less obvious future extension. Remember that this procedure
is used to derive feedback in terms of a hint for the user in cases when he/she is faced with a valid sequent (marked green) but is clueless 
as to which rule to apply next. Equally, the user should receive feedback when he/she uses a rule that does not lead to a completed proof
because it creates an invalid sequent (marked red). Since invalidity is detected via satisfiability of the transformed formula, any model
generated by the underlying SMT solver Z3 can be presented to the user as a counter model. For didactic purposes, Z3's output format would
have to be turned into a suitable presentation of such a counter model, for instance a graphical one. A particular challenge would be 
handling infinite counter models.

However, simply giving the user a counter model to convince him/her that the last step was bad may be a step too big for students who are
struggling with the problem of finding the right proof. It seems like such convincing would have to take smaller steps and require more
effort from the students themselves. This would also be further in line with SCT's design principles; the tool can be seen as a first and 
major step in developing e-learning software for teaching formal reasoning that is based on the well-known didactic principle called 
``Socratic Method/Dialogue''. 

There, learning is understood as a process driven by a teacher asking the right questions. Therefore further developments of the Sequent 
Calculus Trainer will include techniques that resemble the mentioned dialogue. For instance, the software can be extended in a way that
requires the user to present a counter model and then to use game-based model checking techniques \cite{Stirling97} in order to convince 
him/her that it does indeed falsify the sequent at hand.

Further extensions, which are not too surprising, concern the actual implementation. First, there is no need to restrict SCT's power to
provide hints by the abilities of one particular SMT solver. There are others like Yices \cite{Dutertre:cav2014}, CVC4 
\cite{conf/fmcad/DetersR0BT14}, etc. Judging which one is 
best for the purposes here is beyond the scope of this work and also not easy to answer. It is also not necessary as one could run
several SMT solvers in parallel to check sequents for validity and only report an unknown status when all of them have timed out, as seen in KEY (c.f.~Section~\ref{subsec:related_tools}).

Another piece of further work concerning SCT's implementation is a transformation into a browser-runnable application since this would
simplify the running of the tool, and in particular make it more accessible for mobile devices.

\bibliographystyle{eptcs}

\begin{thebibliography}{10}
\providecommand{\bibitemdeclare}[2]{}
\providecommand{\surnamestart}{}
\providecommand{\surnameend}{}
\providecommand{\urlprefix}{Available at }
\providecommand{\url}[1]{\texttt{#1}}
\providecommand{\href}[2]{\texttt{#2}}
\providecommand{\urlalt}[2]{\href{#1}{#2}}
\providecommand{\doi}[1]{doi:\urlalt{http://dx.doi.org/#1}{#1}}
\providecommand{\bibinfo}[2]{#2}

\bibitemdeclare{misc}{ACMIEEERec13}
\bibitem{ACMIEEERec13}
\bibinfo{author}{\surnamestart ACM/IEEE\surnameend} (\bibinfo{year}{2013}):
  \emph{\bibinfo{title}{Computer Science Curricula 2013 --- Curriculum
  Guidelines for Undergraduate Degree Programs in Computer Science}}.
\newblock
  \urlprefix\url{https://www.acm.org/binaries/content/assets/education/cs2013_web_final.pdf}.

\bibitemdeclare{article}{Anderson1995}
\bibitem{Anderson1995}
\bibinfo{author}{J.~R. \surnamestart Anderson\surnameend},
  \bibinfo{author}{A.~T. \surnamestart Corbett\surnameend},
  \bibinfo{author}{K.~R. \surnamestart Koedinger\surnameend} \&
  \bibinfo{author}{R.~\surnamestart Pelletier\surnameend}
  (\bibinfo{year}{1995}): \emph{\bibinfo{title}{Cognitive Tutors: Lessons
  learned}}.
\newblock {\sl \bibinfo{journal}{J.\ of the Learning Sciences}}
  \bibinfo{volume}{4}(\bibinfo{number}{2}), pp. \bibinfo{pages}{167--207},
  \doi{10.1207/s15327809jls0402\_2}.

\bibitemdeclare{book}{Baader:1998:TR:280474}
\bibitem{Baader:1998:TR:280474}
\bibinfo{author}{F.~\surnamestart Baader\surnameend} \&
  \bibinfo{author}{T.~\surnamestart Nipkow\surnameend} (\bibinfo{year}{1998}):
  \emph{\bibinfo{title}{Term Rewriting and All That}}.
\newblock \bibinfo{publisher}{Cambridge University Press},
  \bibinfo{address}{New York, NY, USA}, \doi{10.1017/CBO9781139172752}.

\bibitemdeclare{incollection}{series/faia/BarrettSST09}
\bibitem{series/faia/BarrettSST09}
\bibinfo{author}{C.~W. \surnamestart Barrett\surnameend},
  \bibinfo{author}{R.~\surnamestart Sebastiani\surnameend},
  \bibinfo{author}{S.~A. \surnamestart Seshia\surnameend} \&
  \bibinfo{author}{C.~\surnamestart Tinelli\surnameend} (\bibinfo{year}{2009}):
  \emph{\bibinfo{title}{Satisfiability Modulo Theories}}.
\newblock In: {\sl \bibinfo{booktitle}{Handbook of Satisfiability}}, {\sl
  \bibinfo{series}{Frontiers in Artificial Intelligence and Applications}}
  \bibinfo{volume}{185}, \bibinfo{publisher}{IOS Press}, pp.
  \bibinfo{pages}{825--885}, \doi{10.3233/978-1-58603-929-5-825}.

\bibitemdeclare{article}{DBLP:journals/cj/BornatS99}
\bibitem{DBLP:journals/cj/BornatS99}
\bibinfo{author}{R.~\surnamestart Bornat\surnameend} \&
  \bibinfo{author}{B.~\surnamestart Sufrin\surnameend} (\bibinfo{year}{1999}):
  \emph{\bibinfo{title}{Animating Formal Proof at the Surface: The Jape Proof
  Calculator}}.
\newblock {\sl \bibinfo{journal}{Comput. J.}}
  \bibinfo{volume}{42}(\bibinfo{number}{3}), pp. \bibinfo{pages}{177--192},
  \doi{10.1093/comjnl/42.3.177}.

\bibitemdeclare{article}{DP60}
\bibitem{DP60}
\bibinfo{author}{M.~\surnamestart Davis\surnameend} \&
  \bibinfo{author}{H.~\surnamestart Putnam\surnameend} (\bibinfo{year}{1960}):
  \emph{\bibinfo{title}{A Computing Procedure for Quantification Theory}}.
\newblock {\sl \bibinfo{journal}{J. ACM}} \bibinfo{volume}{7}, pp.
  \bibinfo{pages}{201--215}, \doi{10.1145/321033.321034}.

\bibitemdeclare{inproceedings}{DeMoura:2008:ZES:1792734.1792766}
\bibitem{DeMoura:2008:ZES:1792734.1792766}
\bibinfo{author}{L.~\surnamestart De~Moura\surnameend} \&
  \bibinfo{author}{N.~\surnamestart Bj{\o}rner\surnameend}
  (\bibinfo{year}{2008}): \emph{\bibinfo{title}{Z3: An Efficient SMT Solver}}.
\newblock In: {\sl \bibinfo{booktitle}{Proc.\ 14th Int.\ Conf.\ on Tools and
  Algorithms for the Construction and Analysis of Systems, {TACAS'08}}},
  \bibinfo{series}{LNCS}, \bibinfo{publisher}{Springer}, pp.
  \bibinfo{pages}{337--340}, \doi{10.1007/978-3-540-78800-3\_24}.

\bibitemdeclare{inproceedings}{conf/fmcad/DetersR0BT14}
\bibitem{conf/fmcad/DetersR0BT14}
\bibinfo{author}{M.~\surnamestart Deters\surnameend},
  \bibinfo{author}{A.~\surnamestart Reynolds\surnameend},
  \bibinfo{author}{T.~\surnamestart King\surnameend}, \bibinfo{author}{C.~W.
  \surnamestart Barrett\surnameend} \& \bibinfo{author}{C.~\surnamestart
  Tinelli\surnameend} (\bibinfo{year}{2014}): \emph{\bibinfo{title}{A tour of
  {CVC4}: How it works, and how to use it}}.
\newblock In: {\sl \bibinfo{booktitle}{Proc.\ 14th Int.\ Conf.\ in Formal
  Methods in Computer-Aided Design, {FMCAD'14}}}, \bibinfo{publisher}{IEEE},
  p.~\bibinfo{pages}{7}, \doi{10.1109/FMCAD.2014.6987586}.

\bibitemdeclare{inproceedings}{Dutertre:cav2014}
\bibitem{Dutertre:cav2014}
\bibinfo{author}{B.~\surnamestart Dutertre\surnameend} (\bibinfo{year}{2014}):
  \emph{\bibinfo{title}{Yices 2.2}}.
\newblock In: {\sl \bibinfo{booktitle}{Proc.\ 26th Int.\ Conf.\ on Computer
  Aided Verification, {CAV'14}}}, {\sl \bibinfo{series}{LNCS}}
  \bibinfo{volume}{8559}, \bibinfo{publisher}{Springer}, pp.
  \bibinfo{pages}{737--744}, \doi{10.1007/978-3-319-08867-9\_49}.

\bibitemdeclare{book}{eft_logic_book84}
\bibitem{eft_logic_book84}
\bibinfo{author}{H.-D. \surnamestart Ebbinghaus\surnameend},
  \bibinfo{author}{J.~\surnamestart Flum\surnameend} \&
  \bibinfo{author}{W.~\surnamestart Thomas\surnameend} (\bibinfo{year}{1994}):
  \emph{\bibinfo{title}{Mathematical Logic}}, \bibinfo{edition}{2nd} edition.
\newblock \bibinfo{series}{Undergraduate Texts in Mathematics},
  \bibinfo{publisher}{Springer}, \bibinfo{address}{Berlin},
  \doi{10.1007/978-1-4757-2355-7}.

\bibitemdeclare{mastersthesis}{Ehle17}
\bibitem{Ehle17}
\bibinfo{author}{A.~\surnamestart Ehle\surnameend} (\bibinfo{year}{2017}):
  \emph{\bibinfo{title}{Proof Search in the Sequent Calculus for First-Order
  Logic with Equality}}.
\newblock Master's thesis, \bibinfo{school}{Universit\"{a}t Kassel}.
\newblock \bibinfo{note}{Available via
  \url{https://www.uni-kassel.de/eecs/?id=46992}}.

\bibitemdeclare{inproceedings}{EHL:TTL15}
\bibitem{EHL:TTL15}
\bibinfo{author}{A.~\surnamestart Ehle\surnameend},
  \bibinfo{author}{N.~\surnamestart Hundeshagen\surnameend} \&
  \bibinfo{author}{M.~\surnamestart Lange\surnameend} (\bibinfo{year}{2015}):
  \emph{\bibinfo{title}{The Sequent Calculus Trainer - Helping Students to
  Correctly Construct Proofs}}.
\newblock In: {\sl \bibinfo{booktitle}{Proc.\ 4th Int.\ Conf.\ on Tools for
  Teaching Logic, {TTL'15}}}, pp. \bibinfo{pages}{35--44}.
\newblock \bibinfo{note}{Available via \url{http://arxiv.org/abs/1507.03666}}.

\bibitemdeclare{inproceedings}{DBLP:conf/ticttl/GasquetSS11a}
\bibitem{DBLP:conf/ticttl/GasquetSS11a}
\bibinfo{author}{O.~\surnamestart Gasquet\surnameend},
  \bibinfo{author}{F.~\surnamestart Schwarzentruber\surnameend} \&
  \bibinfo{author}{M.~\surnamestart Strecker\surnameend}
  (\bibinfo{year}{2011}): \emph{\bibinfo{title}{Panda: {A} Proof Assistant in
  Natural Deduction for All. {A} {G}entzen Style Proof Assistant for
  Undergraduate Students}}.
\newblock In: {\sl \bibinfo{booktitle}{Proc.\ 3rd Int.\ Congress on Tools for
  Teaching Logic, {TICTTL 2011}}}, {\sl \bibinfo{series}{LNCS}}
  \bibinfo{volume}{6680}, \bibinfo{publisher}{Springer}, pp.
  \bibinfo{pages}{85--92}, \doi{10.1007/978-3-642-21350-2\_11}.

\bibitemdeclare{article}{Gentzen35a}
\bibitem{Gentzen35a}
\bibinfo{author}{G.~\surnamestart Gentzen\surnameend} (\bibinfo{year}{1935}):
  \emph{\bibinfo{title}{Untersuchungen {\"u}ber das {L}ogische {S}chliessen
  I}}.
\newblock {\sl \bibinfo{journal}{Mathematische Zeitschrift}}
  \bibinfo{volume}{39}, pp. \bibinfo{pages}{176--210},
  \doi{10.1007/BF01201353}.

\bibitemdeclare{phdthesis}{Goedel:1930}
\bibitem{Goedel:1930}
\bibinfo{author}{K.~\surnamestart G{\"o}del\surnameend} (\bibinfo{year}{1930}):
  \emph{\bibinfo{title}{{\"U}ber die {V}ollst{\"a}ndigkeit des
  {L}ogikkalk{\"u}ls}}.
\newblock Ph.D. thesis, \bibinfo{school}{University of Vienna}.

\bibitemdeclare{misc}{GIEmpf16}
\bibitem{GIEmpf16}
\bibinfo{author}{Gesellschaft \surnamestart f.\ Informatik~e.V.\surnameend}
  (\bibinfo{year}{2016}): \emph{\bibinfo{title}{{E}mpfehlungen f{\"u}r
  {B}achelor- und {M}asterprogramme im {S}tudienfach {I}nformatik an
  {H}ochschulen}}.
\newblock
  \urlprefix\url{https://www.gi.de/fileadmin/redaktion/empfehlungen/GI-Empfehlungen_Bachelor-Master-Informatik2016.pdf}.

\bibitemdeclare{article}{Jaskowski:1934}
\bibitem{Jaskowski:1934}
\bibinfo{author}{S.~\surnamestart J{\'a}skowski\surnameend}
  (\bibinfo{year}{1934}): \emph{\bibinfo{title}{On the rules of suppositions in
  formal logic}}.
\newblock {\sl \bibinfo{journal}{Studia Logica}} \bibinfo{volume}{1}, pp.
  \bibinfo{pages}{5--32}.

\bibitemdeclare{article}{LPT2000}
\bibitem{LPT2000}
\bibinfo{author}{M.~J. \surnamestart Lage\surnameend}, \bibinfo{author}{G.~J.
  \surnamestart Platt\surnameend} \& \bibinfo{author}{M.~\surnamestart
  Treglia\surnameend} (\bibinfo{year}{2000}): \emph{\bibinfo{title}{Inverting
  the Classroom: A Gateway to Creating an Inclusive Learning Environment}}.
\newblock {\sl \bibinfo{journal}{J.\ of Economic Education}}
  \bibinfo{volume}{31}(\bibinfo{number}{1}), pp. \bibinfo{pages}{pp. 30--43},
  \doi{10.2307/1183338}.

\bibitemdeclare{article}{Robinson65}
\bibitem{Robinson65}
\bibinfo{author}{J.~A. \surnamestart Robinson\surnameend}
  (\bibinfo{year}{1965}): \emph{\bibinfo{title}{Machine-oriented logic based on
  resolution principle}}.
\newblock {\sl \bibinfo{journal}{Journal of the ACM}} \bibinfo{volume}{12}, pp.
  \bibinfo{pages}{23--41}, \doi{10.1145/321250.321253}.

\bibitemdeclare{article}{DBLP:journals/ijpp/ShneidermanM79}
\bibitem{DBLP:journals/ijpp/ShneidermanM79}
\bibinfo{author}{B.~\surnamestart Shneiderman\surnameend} \&
  \bibinfo{author}{R.~E. \surnamestart Mayer\surnameend}
  (\bibinfo{year}{1979}): \emph{\bibinfo{title}{Syntactic/semantic interactions
  in programmer behavior: {A} model and experimental results}}.
\newblock {\sl \bibinfo{journal}{Int.\ J.\ of Parallel Programming}}
  \bibinfo{volume}{8}(\bibinfo{number}{3}), pp. \bibinfo{pages}{219--238},
  \doi{10.1007/BF00977789}.

\bibitemdeclare{misc}{Stirling97}
\bibitem{Stirling97}
\bibinfo{author}{C.~\surnamestart Stirling\surnameend} (\bibinfo{year}{1997}):
  \emph{\bibinfo{title}{Games for bisimulation and model checking}}.
\newblock \bibinfo{note}{Notes for Mathfit instructional meeting on games and
  computation, Edinburgh}.

\bibitemdeclare{book}{Szabo1969}
\bibitem{Szabo1969}
\bibinfo{editor}{M.~E. \surnamestart Szabo\surnameend}, editor
  (\bibinfo{year}{1969}): \emph{\bibinfo{title}{The Collected Papers of Gerhard
  Gentzen}}.
\newblock \bibinfo{series}{Studies in Logic and The Foundations of
  Mathematics}, \bibinfo{publisher}{North-Holland Publishing Company},
  \doi{10.2307/2272429}.

\bibitemdeclare{article}{Weber2004}
\bibitem{Weber2004}
\bibinfo{author}{K.~\surnamestart Weber\surnameend} \&
  \bibinfo{author}{L.~\surnamestart Alcock\surnameend} (\bibinfo{year}{2004}):
  \emph{\bibinfo{title}{Semantic and Syntactic Proof Productions}}.
\newblock {\sl \bibinfo{journal}{Educational Studies in Mathematics}}
  \bibinfo{volume}{56}(\bibinfo{number}{2}), pp. \bibinfo{pages}{209--234},
  \doi{10.1023/B:EDUC.0000040410.57253.a1}.

\end{thebibliography}

\end{document}